# Ge incorporation inside 4H-SiC during Homoepitaxial growth by chemical vapor deposition.

*Kassem Alassaad[1\*], Véronique Soulière[1], François Cauwet[1], Hervé Peyre[2], Davy Carole[1], Pawel Kwasnicki[2], Sandrine Juillaguet[2], Thomas Kups[3], Jörg Pezoldt[4], Gabriel Ferro[1]*

[1]Université Claude Bernard Lyon 1, CNRS, UMR 5615, Laboratoire des Multimatériaux et Interfaces, 43 Bd du 11 Nov. 1918, 69622 Villeurbanne – France

[2]Laboratoire Charles Coulomb, UMR CNRS 5221 Université Montpellier 2 (France)

[3]FG Werkstoffe der Elektrotechnik, Institut für Mikro- und Nanotechnologien, TU Ilmenau, P.O. Box 100565, 98684 Ilmenau (Germany)

[4]FG Nanotechnologie, Institut für Mikro- und Nanotechnologien, TU Ilmenau, P.O. Box 100565, 98684 Ilmenau (Germany)

**Abstract**. In this work, we report on the addition of GeH$_4$ gas during homoepitaxial growth of 4H-SiC by chemical vapour deposition. Ge introduction does not affect dramatically the surface morphology and defect density though it is accompanied with Ge droplets accumulation at the surface. The Ge incorporation level inside the 4H-SiC matrix, ranging from few $10^{17}$ to few $10^{18}$ at.cm$^{-3}$, was found to be mainly affected by the growth temperature and GeH$_4$ flux. Other growth parameters like C/Si ratio, polarity, or off-orientation did not show any significant influence. On the other hand, adding GeH$_4$ led to the increase of the intentional n type doping level by a factor of 2 to 5 depending on the C/Si ratio in the gas phase.

*Corresponding author*

**Introduction**

The very matured technique for growing SiC epitaxial layers is the chemical vapor deposition (CVD) using classical precursors systems containing C, Si and H elements. Adding a foreign element (except dopants) to the system was widely investigated in the case of Cl, with the well-known improvement in growth rate [1,2]. Beside this, there are very little reports on the investigation of another foreign element to the classical CVD system. One can find few papers on V, Fe, Ti, W, and Nb incorporation [3-7]. Though, in most of the cases the elemental incorporation was either non intentional or in a non-controlled manner. At a first glance, adding an isovalent element like Ge in SiC matrix seems interesting mainly from a fundamental point of view. But some positive surprises are sometimes encountered when searching in different direction.

SiC epitaxy on SiC substrate in the presence of Ge was mostly studied in the case of liquid phase growth. For instance, homoepitaxial 4H-SiC epilayers were grown using Si-Ge melt with the dipping technique [8,9]. These layers were not deeply characterized so that the amount of Ge incorporated inside the 4H lattice is not known. More information on Ge incorporation in SiC matrix can be found in the case of vapor-liquid-solid mechanism (VLS) using Si-Ge melts [10]. It was shown that, in the grown 3C-SiC heteroepitaxial layers, Ge incorporation level could reach up to $1 \times 10^{20}$ at.cm$^{-3}$ [11] but in interstitial site rather than lattice site [12]. The presence of Ge in the 3C-SiC lattice did not seem to affect negatively the photoluminescence properties [13] and these layers even displayed very low interface traps density with SiO$_2$ [14] though this improvement was not attributed to Ge.

In parallel, increased conductivity of 4H-SiC was found after Ge implantation [15,16] but this approach creates a lot of structural damages, which should be avoided. Since Ge is chemically compatible with SiC and is easy to implement via Germane (GeH$_4$) gas, therefore CVD technique can be used for Ge incorporation in order to avoid the implantation drawbacks. As a matter of fact, some of the present authors have recently shown that 4H-SiC epitaxial layers containing Ge have enhanced hall mobility and lower resistivity [17]. This was confirmed by conductive AFM measurements on similar samples that showed increased conductivity in some Ge enriched regions at the surface [18]. So, Ge incorporation in 4H-SiC during epitaxy by CVD seems worth studying not only for fundamental aspects but also for possible material improvement.

We report a complete study related to in situ Ge addition during 4H-SiC homoepitaxy. Effect on growth parameters such as surface morphology, Ge incorporation level and n type doping levels are investigated. Finally, identification of Ge atomic incorporation site inside SiC was determined using atomic location by channeling enhanced microanalysis (ALCHEMI) technique. For more information on this technique and its use for detecting Ge inside SiC, see ref [12].

**Experimental**

The experiments were carried out in homemade epitaxy equipment working at atmospheric pressure. It is composed of a 70 mm diameter vertical cold wall reactor made of quartz and equipped with a homogenization grid on its upper part. The samples were held on a 40 mm diameter cylindrical SiC covered graphite susceptor which was inductively heated. The substrate temperature was monitored with an optical pyrometer connected to a temperature controller. High purity hydrogen (16 slm), silane (2.5 to 5 sccm), propane (2.5 to 8.33 sccm) were used. Germane (GeH$_4$) flux was studied in the 0.01 to 0.1 sccm range. N$_2$ gas (0.25 sccm) was added to some growths in order to study the possible interferences between Ge and this standard n type dopant. The carbon to silicon ratio (C/Si) was varied by changing only the propane flux and keeping the same silane flux. The 4H wafers, from SiCrystal Inc, were 8° off axis (0001) Si and C face and 4° off axis (0001) Si face. For investigating any possible difference in incorporation mechanism between polytypes, a 3C-SiC layer containing low density of twin boundaries, grown by vapor liquid solid mechanism using Si-Sn melt at 1250 °C [7], was also used as seed.

$GeH_4$ was added all along the deposition time together with $N_2$ addition for in situ n-type doping. The growth temperature varied between 1450 to 1600°C. The growth rate and C/Si ratio were changed from 2.3 to 6 µm/h and from 2 to 8 respectively.

The layers were routinely characterized by Nomarski optical microscopy and µ-Raman spectroscopy for surface morphology and polytype determination. The HeNe laser beam ($\lambda$ = 633 nm) was focused down to a spot of a few micrometers squared in a confocal mode configuration. Surface morphology and structure were also investigated by Scanning electron microscope (SEM) and Atomic force microscopy (AFM) Scien-tech (picoscan 5) apparatus in contact mode.

The deposit thickness was deduced from interference fringes of Fourier Transform Infrared Reflectance (FTIR) spectra. The nitrogen doping level was evaluated by mercury microprobe C-V measurements. All Low Temperature Photoluminescence (LTPL) spectra were collected using 30 mW of the 244 nm wavelength of a FreD (Frequency Doubled) $Ar^+$-ion laser as an excitation source. A Triax spectrometer from Jobin-Yvon Horiba fitted with a 600 gr/mm and 2400gr/mm gratings and a cooled CCD camera completed the set-up. The measurements were done at 5K temperature in the wavelength range between 380-700nm. For the Ge incorporation level, secondary ion mass spectrometry (SIMS) technique was implemented. Modified Cameca IMS 4f equipment with an $O^{2+}$ ion source is used. The primary beam voltage was 15 kV and the primary current from 600 – 1200 nA. This resulted in an average etching rate of 30 – 60 Å/s. To calibrate the concentration, Relative Sensitive Factors (RSF) were defined from reference samples where the concentration profiles are well known ($^{74}Ge^+ / ^{13}C^+ = 2.8 \times 10^{18}$). The determined detection limit of $^{74}Ge$ was around few $10^{15}$ at.cm$^{-3}$.

**Results**

*Surface morphology and layer quality*

The surface morphology of the homoepitaxial growth with GeH$_4$ is shown in Fig. 1. On a low magnification scale, it is rather standard with no significant evolution even at the maximum GeH$_4$ flux value (0.1 sccm). Typically, at a growth temperature of 1500 °C and 2.3 μm/h growth rate, the density of triangular and carrot defects on the Ge and non-Ge containing epilayers is found to be similar, in the 400 – 600 cm$^{-2}$ range depending on the C/Si ratio. The only detectable difference when using GeH$_4$ during growth is the presence of dark spots on the surface. When looking closer by SEM, one can see that these spots are in fact droplets, which size and density depend on growth temperature (Fig. 2). Typically, at 1500°C, the average diameter is < 100 nm (Fig. 2a) while it goes up to ~1.5 μm at 1600°C (Fig. 2b).

μ-Raman analyses were performed on sample of Fig. 2b in order to obtain some information about the nature of the droplets and the grown SiC layer itself (Fig. 3). The Raman spectrum recorded several μm away from a droplet was rather standard showing narrow peaks related to the 4H polytype. From the LO$_{964}$ peak, the residual n-type doping was estimated to be at the detection limit, i.e. around or below mid $10^{16}$ cm$^{-3}$ [19]. When performing the μ-Raman analysis on a single droplet, one can see one additional peak located at ~300 cm$^{-1}$ related to Ge-Ge bond and which confirm the presence of pure germanium [20]. The droplets were easily removed chemically by HF-HNO$_3$ and NH$_3$-H$_2$O$_2$ wet etching. AFM was then used to study the surface microstructure. As already observed in a previous work [18], the sample surface presents µm size "depressions" of few nanometers depth. They are probably the fingerprints of the Ge droplets formed during the growth. When looking closer between the droplets, an irregular step and terrace structure is seen (Fig. 4a) whereas the steps are more elongated without GeH$_4$ addition (Fig. 4b). The average step height is calculated to be 7.9 and 9.5 nm respectively for growth with and without GeH$_4$. Despite this difference, the RMS roughness is similar in both case and equal to ~0.4 nm for a 2x2 µm scan.

Low temperature photoluminescence recorded on a typical Ge containing sample of the present study shows that the layer is of high purity (n < 10$^{15}$ cm$^{-3}$), with rather low compensation (Fig. 5). No unknown peak that could be related to the presence of Ge atoms was identified within the energy range.

*Ge incorporation level*

SIMS depth profile on a typical sample grown with GeH$_4$ addition (after droplets wet etching) reveals a homogeneous incorporation level in the entire layer (Fig. 6). The estimated growth rate was 2.3 µm/h which is very similar to the one expected without GeH$_4$ using similar conditions. Ge concentration as a function of GeH$_4$ flow rate is shown in Fig 7. One can notice that the Ge incorporation increases linearly with the precursor flow within the studied range (0.01≤GeH$_4$≤0.1). The maximum achieved level is 5.0x10$^{18}$ cm$^{-3}$. Since the growth rate and C/Si ratio are two essential parameters in SiC epitaxial growth, Ge incorporation dependence as a function of these parameters was studied as shown in Fig. 8 and 9. In the case of C/Si ratio, no clear trend could be detected within the studied range (2-8). However, a moderate increase as a function of growth rate was observed. The Ge incorporation level remains close to 1x10$^{18}$ cm$^{-3}$.

On the other hand, temperature has a more pronounced effect than the two previous parameters as can be seen in Fig. 10. [Ge] decreases of one order of magnitude when growth temperature increases from 1450 to 1600 °C.

Crystallographic dependence of Ge incorporation was studied using different types of substrates (off-orientation, polarity, polytype). These samples were fabricated at 1500 °C using C/Si ratio of 5 and growth rate 2.5 µm/h for 1 hour. The results are summarized in Table 1. One can see that Ge incorporation does not seem to be significantly dependent on the crystallographic aspects.

Finally, when introducing a foreign element in a high purity CVD reactor, an important question is the possible memory effect, i.e. how much unintentional impurity incorporation you can get after a growth using this impurity. This investigation was performed in two ways: 1) a typical "witness" 4H-SiC epilayer was grown after several $GeH_4$ containing growth runs; 2) $GeH_4$ was only added during 10 min in the reactor at 1500°C (to form a network of Ge droplets on the seed surface) before starting a typical CVD growth without $GeH_4$. The conditions for the typical CVD growths are 1500°C with 3 µm/h and C/Si = 5 for 1 hour). In the first case, SIMS analysis pointed out that the amount of Ge in the witness layer is less than or equal to the apparatus detection limit which is $\sim 1 \times 10^{15}$ at/cm$^3$ (figure not shown). In the second case, SIMS analysis showed the presence of Ge in the layer but only located near the epilayer/substrate interface and restricted to several tens of nm only (Fig. 11). In conclusion, there is no significant memory effect due to the use of Ge impurity in our cold wall reactor.

*Effect of $GeH_4$ addition on nitrogen doping*

Since nitrogen doping is a very common step in SiC technology, the possible effect of GeH$_4$ addition on N incorporation was studied. For this set of experiment, N doping was performed using 0.25 sccm of N$_2$ at growth temperature of 1500 °C for 1 hour using constant growth rate of 2.5 µm/h. The results obtained as function of C/Si ratio and GeH$_4$ flux are given in Fig. 12 and 13 respectively. From Fig. 12, one can see that the net n type doping follows the N site competition rule on Si face (N incorporates in C site so that [N] decreases for increasing C/Si ratio) with or without the presence of Ge atoms. But less expected is the fact that the presence of GeH$_4$ increases n doping by a factor from 2 to 5 depending on the C/Si ratio. On the other hand, increasing GeH$_4$ flux does not seem to significantly affect n doping level (Fig. 13).

**Discussion**

*Ge impact on layer quality*

The density of surface defects was shown not to be affected by GeH$_4$ addition. The large majority of these defects have similar length. Taking into account the crystal off-orientation, we postulate that they have been formed at the epilayer/substrate interface [21]. They are most probably caused by substrate imperfection or improper surface preparation, but not by the presence of Ge atoms. The absence of layer quality degradation was confirmed by LTPL characterization. Since no other work has been ever performed on Ge element addition during 4H-SiC homoepitaxy by CVD, we shall compare with studies using other impurities. For instance, it was shown that Al incorporation degrades SiC crystalline quality when incorporated above $10^{20}$ at.cm$^{-3}$ [22] while in the case of N the degradation starts a bit earlier in the $10^{19}$– $10^{20}$ at.cm$^{-3}$ range [23,24]. In the case of "non-dopant" impurities, VCl$_4$ addition to the CVD system leads to surface defect generation for [V] incorporated levels from $1 \times 10^{17}$ at.cm$^{-3}$ and above [3].

In our case, even in the highest Ge containing layer ($5\times10^{18}$ at.cm$^{-3}$), we did not observe any detectable change in surface defect density. Thus, Ge impurity is more like Al or N and does not generate easily crystal defects during growth of 4H-SiC. One can still expect some strain because of Ge presence and its replacement of Si atom [25]. However, high resolution XRD is needed to confirm this hypothesis. On the other hand, Ge accumulates easily on the surface by forming pure Ge droplets, which tend to get bigger when increasing either deposition temperature and/or time. In fact, the main negative impact detected so far for GeH$_4$ addition is on the modification of the surface morphology, either for the step and terrace structure as seen in Fig. 4 or the droplet fingerprints.

*Ge incorporation mechanism*

It was shown in a previous study that, when growing 3C-SiC layers using VLS mechanism from a Si-Ge melt, Ge is incorporating in the layer through nanoscale interstitial clusters, which are detectable, by µ-Raman spectrometry [11,26]. This is not the case in the present study since no Ge-related signal was detected by µ-Raman analyses, even for the case of 3C-SiC polytype (sample E796b of Table 1). In addition, cross sectional TEM analysis performed on one of the epitaxial layers grown with Ge confirms the absence of either Ge inclusions and or Si-Ge-(C) nanoclusters inside the layer [27]. Though Ge may incorporate as isolated interstitial (which will not be detected by Raman spectroscopy or TEM), we will consider here its incorporation on lattice site in SiC matrix. It is worth mentioning here that, according to its covalent radius of 0.122 nm, Ge should rather incorporate on Si site (covalent radius of 0.111 nm) than on C site (covalent radius of 0.077 nm). This was experimentally confirmed in the case of MBE growth of Si-rich cubic SiGeC alloys [28]. In order to clarify the incorporation site of Ge atoms in our experimental conditions, ALCHEMI characterization was performed. By comparing the simulation and experimental results, one comes to the conclusion that Ge atoms are located on Si site, and not on C or interstitial sites (figure 14). This finding supports the theoretical prediction obtained using an anharmonic Keating model showing that substitutional Ge is preferable incorporated on silicon side [C. Guedj, J. Kolodzey, Appl. Phys. Lett. **1999**, 74, 691-693]. As a result, for the rest of the paper, we assume that ~~that~~ Si incorporates on Si site during our growth experiments.

The presence of Ge droplets on the samples after growth is clear evidence that Ge is not massively incorporating inside the lattice, and the excess of Ge brought by the gas phase accumulates on the surface. While this accumulation is time dependent (droplets get bigger with time), none of the SIMS profiles displays any depth evolution of the Ge incorporation level (Fig. 6). This is surprising since these droplets could act as an in-situ extra source of Ge so that one could reasonably expect an increase of Ge incorporation with time. This is not the case and it suggests that, within the studied conditions, Ge is mainly incorporating from the initial $GeH_4$ vapor phase. This is also confirmed by the increase of Ge incorporation with $GeH_4$ flux (Fig. 7). Taking the assumption that Ge droplet formation does not affect significantly Ge incorporation, we can now try to compare Ge incorporation behavior to the ones of some better known impurities (dopants) like Al, N or B (which are not known to accumulate on the surface), with more parallel focus on Al element which is supposed to incorporate on Si site like Ge.

Concerning Ge concentration increase with $GeH_4$ flux, this is of course a classical trend for the dopants [29,30]. The general trend with temperature (Ge incorporation decrease with temperature increase) is also found for the dopants [22,31]. This is explained by an increase of adatoms desorption from the growing surface at higher temperature. From the slope of Fig. 10, one can calculate the activation energy of Ge desorption using Arrhenius law. It is found to be ~ -137 kcal/mol. In the case of Al impurities, the activation energy of desorption which can be found in the literature are of the same order of magnitude though it is systematically higher, ranging from -160 kcal/mol (calculated from the results in ref 22) to -204 kcal/mol [32]. This is to be correlated to the higher vapor pressure of this element compared to Ge. Another similarity with Al impurity is the effect of growth rate. Indeed, Ge incorporation was found to increase with growth rate, like Al does [33]. This can be attributed to the reduced residence time on the surface (and thus reduced desorption rate) of the adatoms at higher growth rates.

Except the previous cited parameters (temperature, Ge flux and growth rate), the other parameters investigated in this study (C/Si ratio and seed crystallography) do not show any significant influence on Ge incorporation level. This is rather uncommon since, for the cases of Al, B or N impurities, these parameters have visible and known influence. For instance, Al atoms, which incorporate on Si site, as should Ge, incorporate less at low C/Si ratio (due to site competition effect) [22]. The simplest explanation would be that Ge incorporates in fact on both C and Si sites. However, this goes against the ALCHEMI results of figure 14 and theoretical predictions obtained in [C. Guedj, J. Kolodzey, Appl. Phys. Lett. **1999**, 74, 691-693]. In addition, as reported in Figure 12, the presence of Ge has a positive effect on N incorporation. Since N is well known to incorporate on C site, if Ge was doing the same then one would expect a competition between N and Ge and thus a reduction of N incorporation when $GeH_4$ is added. This is the reverse which is observed so that we exclude Ge incorporation on C site.

This peculiar insensitivity to C/Si ratio is difficult to explain though one can suggest that it may be linked to our particular Ge incorporation conditions from a Ge-saturated vapor phase and surface. Indeed, in other works, the formation of dopant aggregates or droplets on the growing SiC surface was never reported, even for high concentration of impurity precursors [22,34]. Note that in both cases of B and Al, higher incorporation levels than in the present study can be reached without any surface accumulation.

Concerning the crystallographic aspect, Ge incorporation is found independent on off-orientation, polarity and polytype. In terms of off-orientation, this trend is more or less similar to the ones reported for Al and N incorporation [35], while the polarity is known to have a much higher impact for these dopants. In the present case of Ge impurity, the polarity insensitivity may come from the same reason as for the C/Si ratio insensitivity, i.e. the Ge-saturated vapor phase and surface conditions.

*Ge effect on n type doping*

We will assume for this discussion that the compensation level is low (as seen from LTPL) and that the N atoms incorporated are 100 % activated, so that [N] = n type doping level.

It is shown in Fig. 12 that the presence of $GeH_4$ during $N_2$ in situ doping increases N incorporation, while the amount of $GeH_4$ does not play a significant role in this interaction, as shown in Fig. 13. Since Ge incorporation is supposed to occur on Si site and N one on C site, one may not expect any direct interaction between Ge and N impurities. Furthermore, gas phase interactions are very unlikely when considering the high dilution of $GeH_4$ and $N_2$ precursors. So, interaction at the surface is more probable. One may propose that the presence of Ge atoms on the surface decreases N desorption rate. And if this mechanism is true, it shall be effective at the step edges where the atoms are incorporated. For deeper understanding and explanation of all the aformentioned results, structrural and electrical characterization are in progress.

**Conclusion**

Adding Ge element to the Si-C-H chemical system during 4H-SiC epitaxial growth allowed exploring some growth related fundamental aspects. Its accumulation on the surface, under the form of droplets, does not affect significantly the layer quality or its incorporation level into SiC, which is constant inside the layer. More surprising are the observed deviation from the well-known site competition rule and the interaction with N impurities. More works is still to be done for understanding these findings.

**Acknowledgment**

This work is financially supported by Marie Curie Actions under the project n°264613-NetFISiC.

**Figure captions**

**Fig. 1.** Surface morphology of Ge doped layer grown <u>on 8° off axis seed</u> using 0.02 sccm GeH$_4$ flow at 1550 °C with C/Si ratio of 5 and 2.3 µm/h growth rate for 1 hour.

**Fig. 2.** SEM images of the spherical droplets formed at the surface when introducing GeH$_4$ gas for growth temperature of a) 1500 °C, b) 1600 °C. <u>Both layers were grown on 8° off axis seed using 0.02 sccm GeH$_4$ flow with C/Si ratio of 3.5 and 6 µm/h growth rate for 1 hour.</u>

**Fig. 3.** µ-Raman spectra collected on two areas shown in Fig.2b.

**Fig. 4.** a) AFM surface morphology after wet etching of samples <u>fabricated on 8° off axis (0001) Si face</u>; zoom (2x2 µm$^2$) between the droplets showing the irregular step structure. In b) is shown a 2x2 µm$^2$ scan of a sample grown in similar conditions as in a) but without GeH$_4$ addition. Both samples were grown at 1500 °C having thickness of 2.3 µm.

**Fig. 5.** PL spectrum collected on typical Ge containing sample, grown at 1500 °C using 0.02 sccm of GeH$_4$ and C/Si ratio of 3.5 <u>for 2 hours</u> with 6 µm/h growth rate in the energy range 3.26 - 3.05 eV with marked characteristic phonon position.

**Fig. 6.** SIMS depth profile of sample grown at 1500 °C with 0.02 sccm GeH$_4$ flux for 1h. N concentration profile is shown for localizing the epilayer/substrate interface (note that the N doping level of the layer is below the detection limit of the SIMS apparatus).

**Fig. 7**. Ge incorporation levels as function of GeH$_4$ flow rate. The growth was performed at 1500 °C using a C/Si ratio of 5 with a growth rate of 6 µm/h for 1 hour.

**Fig. 8.** Ge incorporation level as a function of C/Si ratio. Samples were grown at 1500 ˚C using 0.02 sccm of $GeH_4$ and fixing the silane flux to 5 sccm (growth rate is 6 um/h and growth time is 1 h).

**Fig. 9.** Ge incorporation level as a function of growth rate. Samples were grown at 1500 ˚C using 0.02 sccm of $GeH_4$ with a fixed C/Si ratio of 5 (growth rate is 6 um/h and growth time is 1 hour)

**Fig. 10.** Ge incorporation as a function of temperature. Samples were grown using fixed $GeH_4$ flux of 0.02 sccm and C/Si ratio of 3.5 (growth rate is 6 um/h and growth time is 1 hour)

**Fig. 11.** SIMS profile collected from a sample grown without $GeH_4$ during the deposition time but with 0.1 sccm $GeH_4$ deposition for 10 minutes prior to the growth at 1500 ˚C.

**Fig. 12.** N type doping level as a function of C/Si ratio for layers grown with or without $GeH_4$.

**Fig. 13.** N type doping level as a function of injected $GeH_4$ flux. Samples were grown at 1500 ˚C using fixed growth rate and C/Si ratio of 5.

**Table caption**

**Table 1.** Ge concentration inside samples grown using the same conditions but on various seeds.

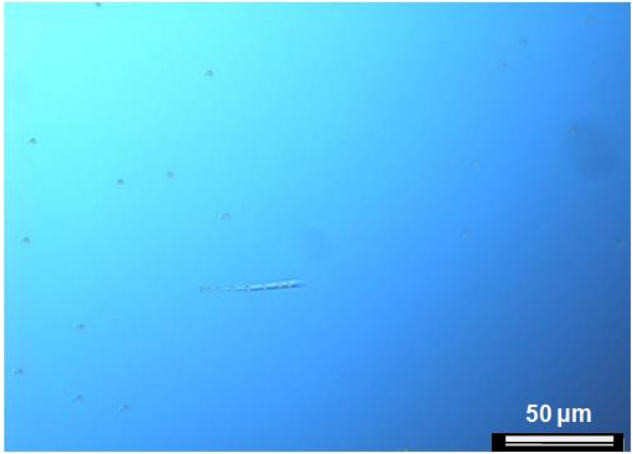

Fig. 1. Surface morphology of Ge doped layer grown on 8° off axis seed using 0.02 sccm GeH$_4$ flow at 1550 ˚C with C/Si ratio of 5 and 2.3 µm/h growth rate for 1 hour

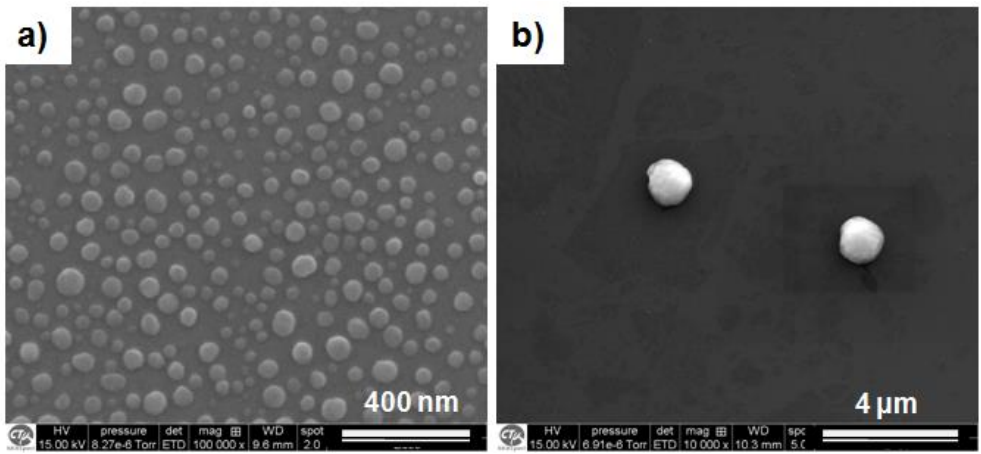

Fig. 2. SEM images of the spherical droplets formed at the surface when introducing GeH$_4$ gas for growth temperature of a) 1500 °C, b) 1600 °C. Both layers were grown on 8° off axis seed using 0.02 sccm GeH$_4$ flow with C/Si ratio of 3.5 and 6 µm/h growth rate for 1 hour.

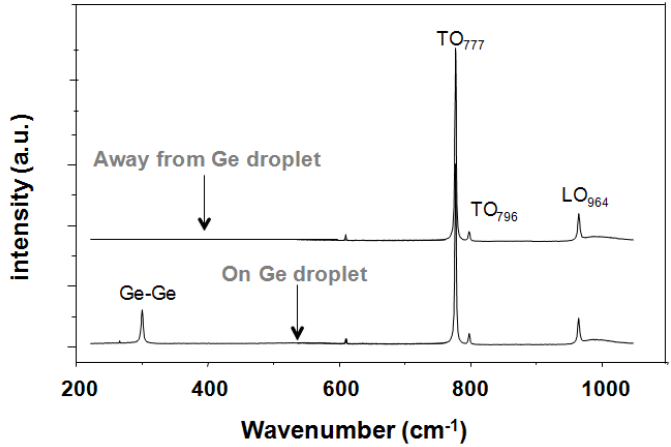

Fig. 3. μ-Raman spectra collected on two areas shown in Fig.2b.

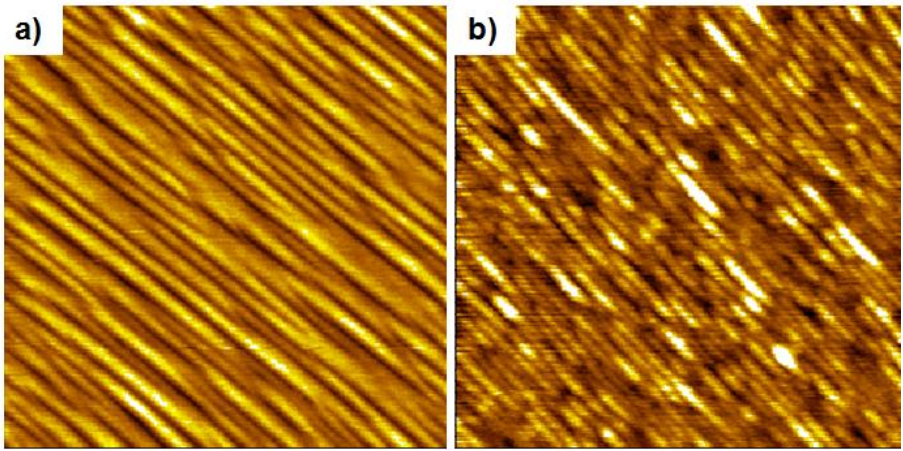

Fig. 4. a) AFM surface morphology after wet etching of samples fabricated on 8° off axis (0001) Si face; zoom (2x2 μm$^2$) between the droplets showing the irregular step structure. In b) is shown a 2x2 μm$^2$ scan of a sample grown in similar conditions as in a) but without GeH$_4$ addition.

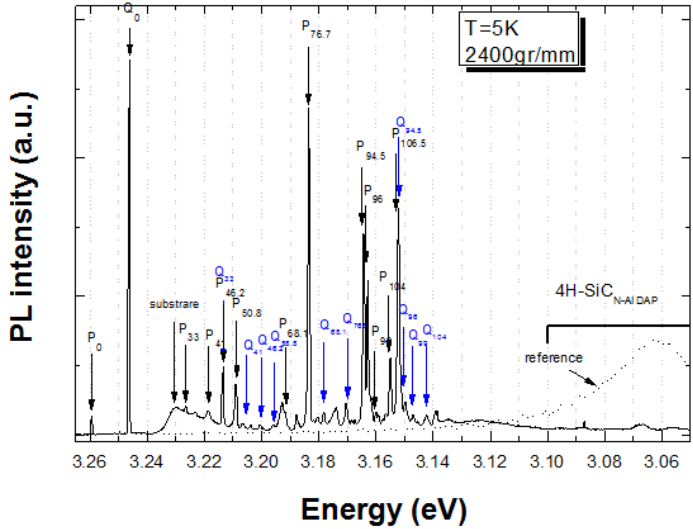

Fig. 5. PL spectrum collected on typical Ge containing sample, grown at 1500 °C using 0.02 sccm of GeH$_4$ and C/Si ratio of 3.5 for 2 hours with 6 μm/h growth rate in the energy range 3.26 - 3.05 eV with marked characteristic phonon position.

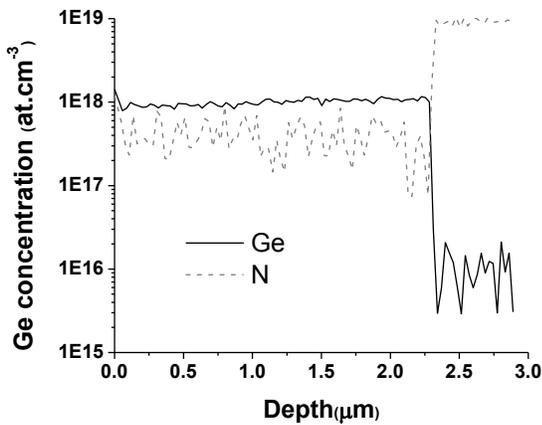

Fig. 6. SIMS depth profile of sample grown at 1500 ˚C with 0.02 sccm GeH$_4$ flux for 1h. N concentration profile is shown for localizing the epilayer/substrate interface (note that the N doping level of the layer is below the detection limit of the SIMS apparatus).

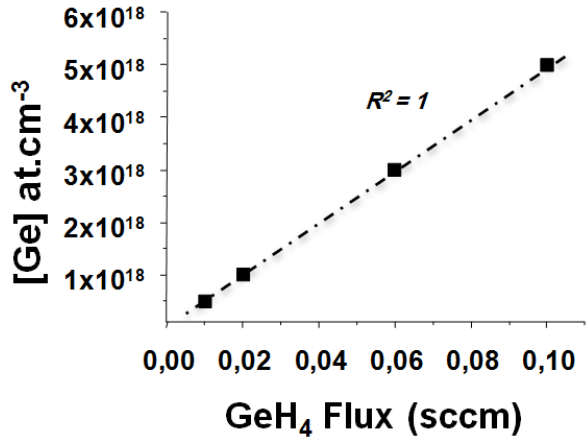

Fig. 7. Ge incorporation levels as function of GeH$_4$ flow rate. The growth was performed at 1500 °C using a C/Si ratio of 5 with a growth rate of 6 µm/h for 1 hour.

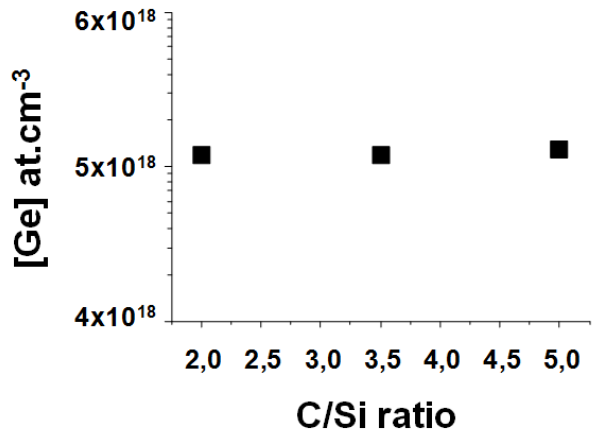

Fig. 8. Ge incorporation level as a function of C/Si ratio. Samples were grown at 1500 °C using 0.02 sccm of GeH$_4$ and fixing the silane flux to 5 sccm (growth rate is 6 um/h and growth time is 1 hour)

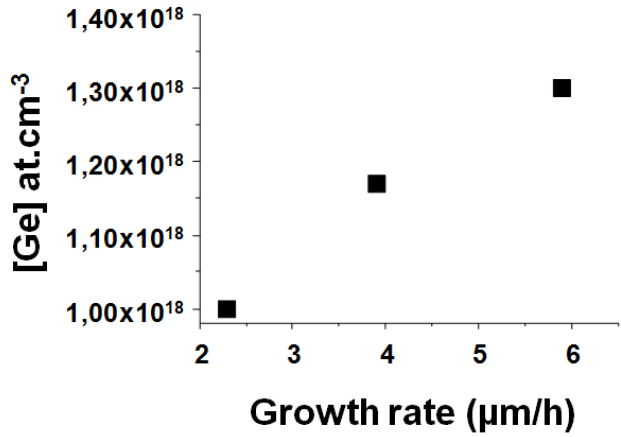

Fig. 9. . Ge incorporation level as a function of growth rate. Samples were grown at 1500 ˚C using 0.02 sccm of GeH$_4$ with a fixed C/Si ratio of 5 (growth rate is 6 um/h and growth time is 1 hour)

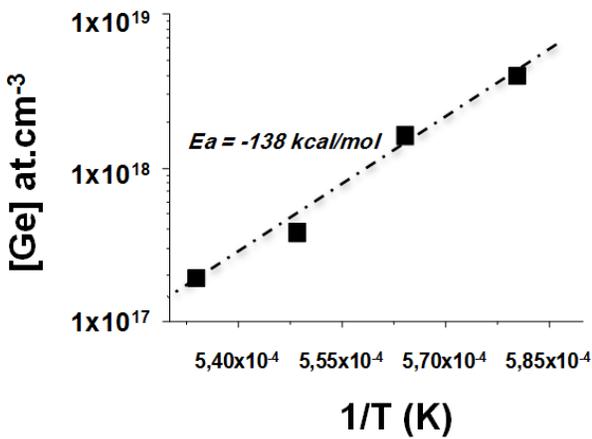

Fig. 10: Ge incorporation as a function of temperature. Samples were grown using fixed GeH$_4$ flux of 0.02 sccm and C/Si ratio of 3.5 (growth rate is 6 um/h and growth time is 1 hour)

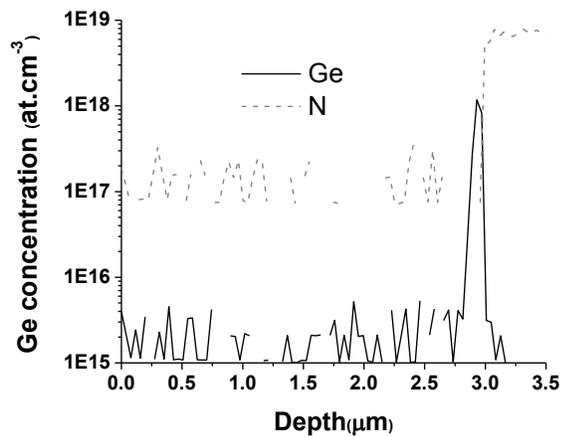

Fig. 11. SIMS profile collected from a sample grown without $GeH_4$ during the deposition time but with 0.1 sccm $GeH_4$ deposition for 10 minutes prior to the growth at 1500 ˚C.

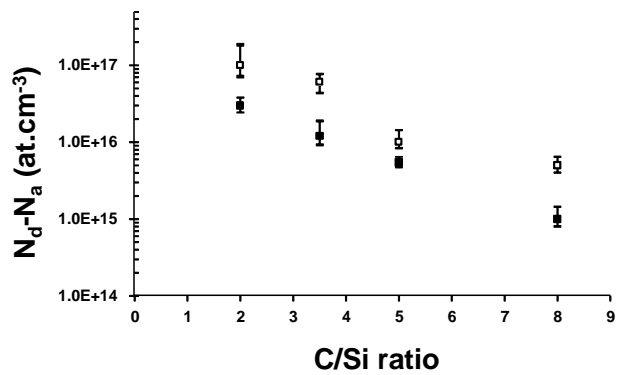

Fig. 12. N type doping level as a function of C/Si ratio for layers grown with or without $GeH_4$.

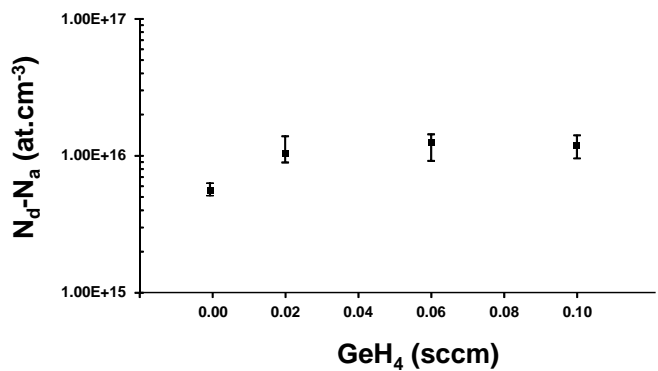

Fig. 13. N type doping level as a function of injected $GeH_4$ flux. Samples were grown at 1500 ˚C using fixed growth rate and C/Si ratio of 5.

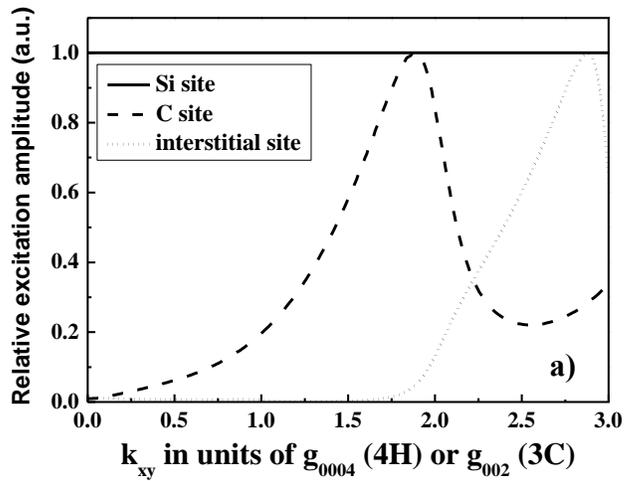

Figure 14. a) ALCHEMI simulation showing Bloch state excitation (normalized to Si-signal) for different atomic sites in SiC material and b) Ge signal evolution on a Ge doped sample grown in this study. Both results are given as a function the reciprocal lattice vector in units of $g_{hkl}$.

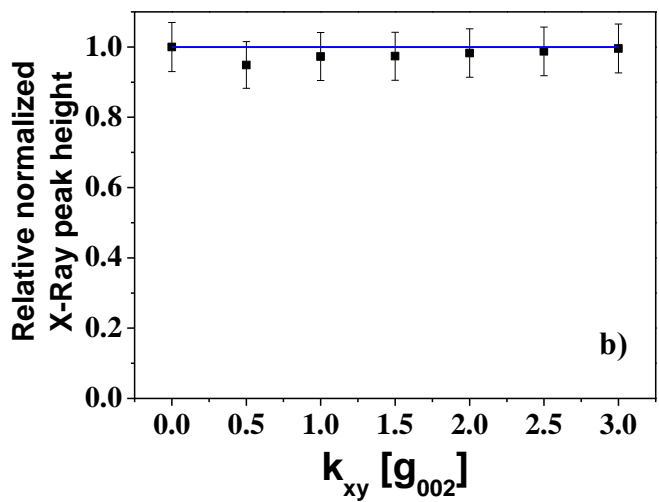

Table 1. Ge concentration inside samples grown using the same conditions but on various seeds.

| Sample Name | Substrate polytype | Off orientation from (0001) | Polarity | [Ge] at.cm$^{-3}$ |
|---|---|---|---|---|
| E813a | 4H-SiC | 8 Degrees | Si face | $9.7 \times 10^{17}$ |
| E813b | 4H-SiC | 8 Degrees | C face | $1.0 \times 10^{18}$ |
| E813c | 4H-SiC | 4 Degrees | Si face | $9.8 \times 10^{17}$ |
| E796b | 3C-SiC | On-axis (111) | Si face | $9.9 \times 10^{17}$ |